\documentclass[sigconf]{acmart}

\AtBeginDocument{%
  }

\setcopyright{none}
\acmDOI{}
\acmISBN{}
\acmConference[LIMITS '26]{12th Workshop on Computing within Limits}{June 23--25, 2026}{Online}

\usepackage{booktabs}
\usepackage{siunitx}
\usepackage{listings}
\usepackage{xcolor}
\usepackage{ifthen}
\usepackage{graphicx}
\usepackage{makecell}
\usepackage{xurl}
\usepackage{pgfplots}
\pgfplotsset{compat=1.18}

\lstdefinelanguage{json}{
  basicstyle=\ttfamily\footnotesize,
  showstringspaces=false,
  breaklines=true,
  morestring=[b]",
  literate=
   *{0}{{0}}{1}
    {1}{{1}}{1}
    {2}{{2}}{1}
    {3}{{3}}{1}
    {4}{{4}}{1}
    {5}{{5}}{1}
    {6}{{6}}{1}
    {7}{{7}}{1}
    {8}{{8}}{1}
    {9}{{9}}{1}
    {:}{{:}}{1}
    {,}{{,}}{1}
    {\{}{{\{}}{1}
    {\}}{{\}}}{1}
    {[}{{[}}{1}
    {]}{{]}}{1},
  morekeywords={true,false,null}
}

\title{Measure Once, Model Everywhere:\\ Model-Based Per-Request Resource Consumption for HTTP}

\author{Geerd-Dietger Hoffmann}
\affiliation{
  \institution{Green Coding Solutions,\\ {\small University of Potsdam, HTW Berlin}}
  \city{Potsdam}
  \country{Germany}
}
\email{didi@green-coding.io}

\author{Verena Majuntke}
\affiliation{
  \institution{HTW Berlin}
  \city{Berlin}
  \country{Germany}}
\email{verena.majuntke@htw-berlin.de}

\begin{document}

\begin{abstract}
Recent proposals for HTTP-based sustainability disclosure focus on \textbf{what} environmental information should be transmitted at the protocol boundary, for example through response headers, but leave open the practical question of \textbf{how} such per-request values can be generated in realistic deployments. This paper addresses that implementation gap. We present a model-based approach for estimating resource consumption and $CO_2e$ per HTTP request without requiring fine-grained production power telemetry. The approach benchmarks endpoints offline under controlled conditions, derives compact endpoint-specific energy models from observable request features, and evaluates these models online at the HTTP server boundary. We implement this mechanism as an nginx extension that loads a JSON model registry and emits per-request metadata for energy, grid intensity, embodied emissions, and total request-level impact. We show that heterogeneous request classes can be represented with constant, linear, and piecewise models, and that the same approach extends to endpoints whose dominant cost driver is only visible at the application layer through inputs such as token counts. Our evaluation indicates that the approach is operationally feasible and introduces only low runtime overhead.
\end{abstract}
\maketitle

\section{Introduction}

Software is increasingly delivered through HTTP interfaces built on opaque
stacks of virtualised infrastructure, managed platforms, and third-party
services. The dominant cloud-native model presents these resources as
elastic and available on demand: compute, storage, and bandwidth can be
acquired through a single API call or a dashboard click, with billing and
quotas as the primary feedback channels. The same abstractions that make
this delivery scalable also conceal the physical and ecological costs of
operating the underlying infrastructure
\cite{mytton2020hiding,parssinen2018onlineadvertising,han2025fairco2}, so
that at the interface where requests are issued the system appears
practically limitless. Users can observe latency and price, but not the
energy use or emissions of a single request. Providers face a similar
challenge: even when disclosure is desired, request execution spans
multiple systems, and fine-grained power telemetry is typically
unavailable.

A recurring principle in sustainable computing is to expose information at the
interface where decisions are made. The Software Carbon Intensity (SCI)
specification expresses emissions relative to a functional unit \(R\)
\cite{gsf-sci}. For web systems, the natural functional unit is the HTTP
request. If request-level impact were available at the protocol boundary,
clients could incorporate it into decisions such as retry behaviour, batching,
or prompt design.

HTTP is a natural disclosure surface, and prior work has proposed response headers for communicating carbon information~\cite{martin-http-carbon-emissions-scope-2-00,httpwg-admin-issue-52-carbon-emissions,ietf-http-wg-mailinglist-carbon-header-thread,besleaga-green-sustainability-header-00}. However, these efforts focus on \emph{what} to disclose, not on how to derive per-request values in practice.

Per-request disclosure does not by itself reduce demand, but it is a precondition for the design strategies that do. Sufficiency- and sobriety-oriented approaches~\cite{nardi2018computingwithinlimits,pargman2014rethinking}, critiques of growth-oriented computing infrastructure~\cite{becker2023insolvent}, and low-impact server designs such as the solar website~\cite{dedecker2018solar} all require the marginal environmental cost of an interaction to be legible at the boundary where the request is issued. Without that legibility, providers and clients cannot distinguish high-cost from low-cost interactions, and ecological cost remains structurally excluded from the design decisions that determine aggregate consumption.

This paper addresses the implementation gap by introducing a practical, model-based method for deriving per-request energy and emissions. Instead of relying on production telemetry, endpoints are benchmarked offline under controlled conditions to derive compact energy models that are then evaluated at request time using variables already observable at the HTTP boundary, such as request duration, transferred bytes, or application-level metadata. HTTP-based disclosure is used as one instantiation but is not required by the method.

This paper makes four contributions:
(1) It separates the protocol-level question of \emph{what to disclose} from the systems question of \emph{how to generate request-level values}, and
identifies the latter as a key gap in current HTTP sustainability discussions.
(2) It proposes a measurement-to-model pipeline that derives compact, interpretable, and executable energy models for HTTP endpoints from controlled
offline benchmarks.
(3) It demonstrates how these models can be evaluated at the HTTP server boundary and exposed as request-level sustainability metadata, while keeping
underlying assumptions explicit.
(4) It evaluates the feasibility, expressiveness, and runtime overhead of the approach across heterogeneous endpoints, including conventional web services
and an inference endpoint that serves as an example of an endpoint requiring application-supplied features.

\section{Related Work}
\label{sec:rw}

Assessing the environmental impact of software requires more than raw
instrumentation. Guldner et al.\ propose the Green Software Measurement Model
(GSMM), which emphasises that sustainability assessment depends on explicit
modelling choices, system boundaries, and the interpretation of measurements
\cite{guldner2024gsmm}. The Software Carbon Intensity (SCI) specification
similarly frames impact relative to a functional unit \(R\) \cite{gsf-sci}.
Together, these approaches highlight that measurement alone is insufficient and
must be complemented by modelling.

Prior work also shows that energy use varies significantly with workload
characteristics. Pärssinen et al.\ quantify the environmental impact of web
interactions \cite{parssinen2018onlineadvertising}, and Sala et al.\ demonstrate
that sustainability metrics can be integrated into operational web monitoring
\cite{sala2024greenwebmeter}. In the context of AI, Desislavov et al.\ and
Poddar et al.\ show that inference energy depends on runtime properties such as
execution time and output length \cite{desislavov2023trends,poddar2025towards}.
These works establish that fine-grained variation exists, but primarily analyse
it offline or at aggregate levels.

Attributing energy and emissions in cloud environments is inherently difficult.
Fair-CO2 highlights the challenge of allocating both operational and embodied
emissions across shared infrastructure and multi-tenant systems
\cite{han2025fairco2}. This limits the availability of directly observable
per-request values in production environments.

Recent work has explored HTTP as a surface for communicating sustainability
information. Martin proposes a "Carbon-Emissions-Scope-2" response header
\cite{martin-http-carbon-emissions-scope-2-00}, and more recent drafts define
structured sustainability headers \cite{besleaga-green-sustainability-header-00}.

A parallel body of work argues that computing should be reconceived
under non-negotiable ecological constraints rather than treated as a
domain to be optimised within a growth-oriented status quo
\cite{nardi2018computingwithinlimits,pargman2014rethinking,becker2023insolvent}.
Several strands of this literature intersect directly with the present
paper. Raghavan and Ma argue that the energetic and material substrate
of the Internet must be made legible to system design rather than treated
as an externality \cite{raghavan2011energy}. Mytton shows that
standardised cloud reporting tends to hide rather than expose the
emissions associated with computation, reinforcing the case for
disclosure mechanisms that operate close to the unit of consumption
\cite{mytton2020hiding}. Constraint-led design has also been demonstrated
empirically: the solar-powered Low-tech Magazine website exposes its
energy budget directly to readers and modulates content under poor
solar conditions, treating available energy as a binding parameter of
service behaviour rather than as an externality
\cite{dedecker2018solar}. The method proposed in this paper is a
direct substrate for these positions. Sufficiency and sobriety
arguments require an interface at which the marginal cost of an
interaction can be seen and compared; degrowth-oriented critiques
require ecological cost to be inescapable at the point of demand
rather than externalised through cloud abstractions; low-impact
server designs (solar websites, low-tech publishing, intermittent
computing) require operating envelopes that the underlying system can
both honour and communicate. The mechanism developed here supplies
that substrate without prescribing a specific level of consumption.

Existing work therefore establishes the need for modelling, demonstrates
workload-level variation, and defines potential disclosure interfaces.
However, it does not provide a practical method for generating per-request
energy and emissions estimates under realistic deployment constraints. In
particular, the step from controlled measurements to request-level values
available at the interface remains insufficiently specified.

\section{Per-Request Energy Estimation}
\label{sec:methods}

This section presents our approach for deriving per-request energy and emissions estimates from controlled measurements under realistic observability constraints. The approach separates offline measurement from online estimation: endpoints are characterised offline under controlled conditions, and the resulting models are evaluated at request time using variables observable at the HTTP boundary.\footnote{Throughout this paper, ``HTTP boundary'' (equivalently, ``HTTP server boundary'') denotes the request/response surface at the reverse proxy; ``machine boundary'' denotes the point of physical energy measurement; ``system boundary'' denotes the accounting scope, discussed in Section~\ref{sec:limitations}.} A method addressing this setting must satisfy four design requirements: it must (i) operate under limited observability, including deployments without fine-grained power sensors; (ii) be efficient enough to evaluate on every request; (iii) remain interpretable, so that providers can communicate the assumptions underlying the reported values; and (iv) capture heterogeneous endpoint behaviour, from near-constant-cost CRUD (create, read, update, delete) requests to endpoints whose cost depends on request-specific inputs only visible at the application layer. A concrete instantiation of the method follows in Sections~\ref{sec:measurement} and~\ref{sec:implementation}.

\subsection{Endpoint energy models}
\label{subsec:model}

Each endpoint \(e\) is associated with a model \(\widehat{E}_e\) that predicts request energy from one or more observable features (e.g., request duration, transferred bytes, HTTP method, response status, request-specific parameters). The feature set is not fixed; it captures quantities that are either already available at the HTTP boundary or can be exposed with minimal application changes. We restrict the model space to a small set of families that balance expressiveness, interpretability, and efficient runtime evaluation. The full collection of fitted models, together with the per-deployment carbon parameters from Section~\ref{subsec:emissions}, is stored as a \emph{registry}: a compact data structure indexed by endpoint identifier that the server loads at runtime (Section~\ref{sec:implementation}).

Three patterns recur in practice: some endpoints show negligible variation across requests, others scale approximately with an observable feature, and some exhibit distinct scaling regimes that no single straight line fits well. We use one model family for each case.

\textbf{Constant model.}
When energy does not vary appreciably with the input, a single
representative value suffices. The endpoint is assigned a fixed
predicted energy \(c_e\), set to the mean energy observed during
calibration:
\[
\widehat{E}_e = c_e.
\]

\textbf{Linear model.}
When energy varies in proportion to an observable quantity, a linear
model captures this scaling as a baseline plus a per-feature
contribution. For a single feature \(x\),
\[
\widehat{E}_e(x) = \beta_0 + \beta_1\, x,
\]
where the intercept \(\beta_0\) is the baseline energy of a request and
the coefficient \(\beta_1\) the additional energy per unit of \(x\).
Further features can be incorporated by adding more
\(\beta_j\, x_j\) terms. Coefficients are fitted to the calibration
data by standard least-squares estimation.

\textbf{Curve model.}
Some endpoints exhibit distinct scaling regimes, for example behaving
as near-constant for small inputs and growing sharply for large ones.
Rather than committing to a particular non-linear functional form, the
curve model represents the endpoint by a small set of measured points,
each pairing a feature value \(z_i\) with the energy \(E_i\) observed
at that value. At request time, the predicted energy for an input
\(z\) is obtained by linear interpolation between the two adjacent
measured points. The model therefore reads as a sequence of
straight-line segments through the calibration data: it is simple to
evaluate while allowing the slope to change between regimes.

\textbf{Application-supplied features.}
Some predictors are not visible at the HTTP server layer but are known to the application or a downstream runtime (e.g., token counts for inference, result-set cardinality, cache-hit indicators). Such quantities can be exposed to the server as request-scoped metadata via internal headers or server variables, and used as additional inputs to any of the three model families above.

\textbf{Model family selection.}
Family assignment is treated as a modelling decision informed by workload-scaling benchmarks: energy is measured under controlled variation of the relevant workload parameter, and the family that best reflects the observed scaling is selected. Fitting then reduces to standard parameter estimation (averaging for constant models, least-squares for linear, retained support points for curves). The method does not preclude automated selection or uncertainty-aware fitting, but the present formulation makes the manual modelling step explicit so that the resulting registry remains inspectable.

\subsection{Operational and embodied emissions}
\label{subsec:emissions}

The models above predict per-request energy. To obtain a sustainability metric that is comparable across deployments, we convert this energy into carbon emissions, distinguishing operational emissions (energy consumed during the request) from embodied emissions (a share of the hardware's manufacturing impact). Reporting both components separately keeps the underlying assumptions visible at the boundary.

\textbf{Operational emissions.}
Operational emissions are obtained by multiplying the predicted energy by the grid carbon intensity \(I\) at the time of execution:
\[
\widehat{C}^{op}_e = \widehat{E}_e \cdot I.
\]
The grid carbon intensity \(I\) is an input to the method, not a fixed
property of it. The same equation supports two regimes. In the
\emph{static} regime, \(I\) is set to a representative value for the
deployment region, for example a published annual or location-specific
average, and the reported operational emissions correspond to that
average. In the \emph{dynamic} regime, \(I\) is refreshed at request
time or on a short interval from an external signal such as a public
carbon-intensity service (for example, Electricity
Maps~\cite{electricitymaps} or WattTime~\cite{watttime}) or an
internal monitoring source; the reported value then reflects the grid
state that applied while the request was served. In both regimes, the
value of \(I\) used for the conversion is itself disclosed alongside
the resulting emissions, so that downstream consumers can recover the
assumption. Units (for example, \(\mathrm{mWh}\) for energy and
\(\mathrm{gCO_2e/kWh}\) for intensity) are reconciled in the
implementation.

\textbf{Embodied emissions.}
Embodied emissions capture the carbon cost of manufacturing and provisioning the hardware. Following a time-based allocation, the total embodied emissions \(M\) of the system are amortised over an assumed operational lifetime \(T\) to give a constant embodied rate \(r^{emb} = M/T\). The share attributed to a request with processing time \(t\) is then
\[
\widehat{C}^{emb}_e = r^{emb} \cdot t.
\]
This is a simple and transparent allocation scheme. Other schemes, such as utilisation-weighted or capacity-based allocation, can be substituted depending on the deployment context~\cite{iso14040,iso14044,ghgproduct,gsf-sci}; the rest of the method does not depend on this choice.

Total per-request emissions are the sum of the two components:
\[
\widehat{C}_e = \widehat{C}^{op}_e + \widehat{C}^{emb}_e.
\]

Any concrete instantiation requires accounting choices, for example whether benchmark labels are derived from total machine energy or from idle-subtracted measurements (where the machine's baseline idle energy is subtracted to isolate the request-dependent signal), whether protocol overhead is included in calibration, and how embodied carbon is allocated. These choices do not affect the structure of the equations above but do affect interpretation, and must therefore be disclosed alongside the reported values. The specific choices made in this paper are stated in Section~\ref{sec:evaluation}; limitations are consolidated in Section~\ref{sec:limitations}.

\section{Implementation}
\label{sec:measurement}
This section instantiates the method of Section~\ref{sec:methods} on a concrete benchmark application. It describes how we generate per-request energy labels \(E\) (in \(\mathrm{mWh}\)) from controlled experiments. These measurements form the basis for fitting the per-endpoint energy models described in Section~\ref{sec:methods}. For each request, we record energy at the machine boundary and associate it with request-level observations obtained during benchmarking, including request processing time, data transfer volume, and endpoint-specific workload parameters.

All measurements were executed using the Green Metrics Tool (GMT) \cite{greenmetricstool} in a controlled measurement setup. Experiments were conducted on a dedicated machine to ensure reproducibility and minimise interference from background workloads.

We used the GMT ``ML/AI Profiling'' profile on a server-class system with the following configuration: an Intel Core i5-9600K @ 3.70\,GHz (6 cores / 6 threads; Turbo enabled; DVFS enabled) with 32\,GB of memory, an NVIDIA GeForce GTX 1080 (8\,GB VRAM, CUDA 12.2, driver 535.274.02), power measurement via MCP39F511N and IPMI (as provided by GMT), running Ubuntu 24.04 (NOP Linux) \cite{greenmetricstool}.

Turbo boost and dynamic frequency scaling were left enabled to reflect typical deployment conditions.
\subsection{Measurement Practice}
To obtain stable energy measurements, we followed the Green Coding best practices for controlled benchmarking \cite{greencoding_bestpractices}. These practices stress that software-energy measurements are sensitive to noise and should therefore be performed under controlled conditions, with temperature effects, idle baselines, and run-to-run variance taken into account.

Concretely, we ran the service under test (SUT) on an isolated machine running NOP Linux, a minimal environment with non-essential processes disabled to reduce measurement noise. Before sampling, we warmed up the SUT to reach a steady state. We then recorded an idle baseline for a comparable interval. During the experiments, we varied one workload factor at a time, such as payload size or token count, and repeated each condition often enough to make the signal distinguishable from measurement noise. As recommended in the Green Coding documentation, the exact number of repetitions depended on the variance and expected signal strength of the endpoint under test.

The whole test suite was executed 10 times. All measurements are available under \url{https://metrics.green-coding.io/runs.html?uri=https%3A%2F%2Fgithub.com%2Fribalba%2FLimits-2026&show_archived=true&show_other_users=true}.

\subsection{Benchmark Application Setup}
We implemented a minimal REST-style ToDo application in Python that serves HTTP
requests using SQLite as the backend database. A user can log in and log out,
create ToDo items with a title, text, and optional file attachment, retrieve
all ToDo items, and mark individual items as done. In addition, the application
provides an LLM endpoint for sentence completion. The application and benchmark
harness are publicly available \cite{easytodo}. While SQLite is a simple
dependency, the GMT harness supports measuring more complex multi-service
deployments. We use this application to cover a range of endpoint behaviours.
The endpoints are described in Table~\ref{tab:endpoints}.

\begin{table}
    \centering
    \begin{tabular}{p{1.7cm} p{6.3cm}}
        /login & Given a username and password returns an authenticated session cookie \\
        /logout & Invalidates the authenticated session \\
        /createToDo  & Takes a title, text and attachment and creates a new "todo" item in the database \\
        /done &  Given a todo id marks it as done in the DB \\
        /getToDos  & Returns all the todos of the user \\
        /deleteAllToDos & Deletes all the todos of the user \\
        /ai & Given a text sends it to an LLM for autocompletion and returns the ai response \\
    \end{tabular}
    \caption{REST endpoints used in the benchmark application}
    \label{tab:endpoints}
\end{table}

HTTPS is excluded from the measurements to avoid entangling endpoint energy with TLS handshake and encryption overhead; the implications of this scoping choice and possible mitigations are discussed in Section~\ref{sec:limitations}.

\subsection{Benchmarking}
Each endpoint was benchmarked with a workload script designed to expose its expected scaling factors. Stateful endpoints were called with a pre-acquired session cookie reused across runs to avoid repeatedly measuring login overhead. Repetition counts are chosen per endpoint: 100 calls per condition for the lightweight REST endpoints (\texttt{/login}, \texttt{/logout}, \texttt{/createToDo}, \texttt{/getToDos}, \texttt{/done}), where single-request energy is too small to resolve above measurement noise; and 10 calls per prompt profile for \texttt{/ai}, where each inference call is long enough to produce a measurable signal on its own. \texttt{/createToDo} conditions vary either text length (100, 1\,000, 10\,000, 100\,000 chars) or attachment size (1\,KB, 10\,KB, 100\,KB, 1\,MB, 5\,MB); \texttt{/getToDos} and \texttt{/done} run against the populated state; \texttt{/deleteAllToDos} cleans up between conditions.

The \texttt{/ai} endpoint serves \texttt{gemma3:1b} via a local \texttt{ollama} container on the same machine. Inference behaviour is stabilised by downloading the model before sampling, prewarming caches, and holding decoding settings constant. We treat prompt-token and generated-token counts as the application-supplied features for \texttt{/ai} (Section~\ref{subsec:model}): the application layer reads them from the inference runtime and exposes them as response-scoped metadata, so a token-linear model can be fitted without GPU telemetry. Prior work supports this choice: under fixed serving conditions, latency and energy scale approximately with output-token count~\cite{yang2024queueing,poddar2025towards,caravaca2025promptstopower}. The inference endpoint is included as an example of a workload whose dominant cost driver is only visible at the application layer; we make no claim about hosted inference services at scale, where a substantial share of emissions occurs outside the reverse-proxy boundary.

\subsection{Per-Request Energy Labels}

We derive per-request energy labels \(E\) (in \(\mathrm{mWh}\)) from machine-level energy measurements collected during benchmarking~\cite{gmt-run-benchmark-energy}. For each condition, we compute:
\[
E = \frac{E_{\text{machine}}}{n},
\]
where \(E_{\text{machine}}\) is the total energy measured during the benchmark interval and \(n\) is the number of requests executed in that interval.

This yields an average per-request estimate that approximates the energy attributable to the endpoint under the given workload condition. The estimate assumes that requests within each condition are homogeneous and that background effects remain stable during the measurement interval.

We verify that energy varies consistently with the controlled workload parameter under input sweeps, indicating that the derived labels capture the intended endpoint behaviour.

\subsection{Runtime-Observable Features}
For each measured request, we associate the energy label \(E\) with features that are available (or can be made available at low cost) at the HTTP server boundary. The features below were selected on two criteria: that they can be measured at the boundary without invasive instrumentation, and that they vary meaningfully across the requests of interest. Headers such as \texttt{Content-Type} are intentionally omitted because in this case study, as is typical for REST-style services, they are effectively fixed per endpoint and method, and therefore do not provide additional discriminating signal beyond what endpoint identity and HTTP method already carry.
\begin{itemize}
  \item \textbf{Processing time \(t\):} request handling time as observed by the reverse proxy (e.g., nginx \texttt{\$request\_time}-style timing variables).
  \item \textbf{Request and Response bytes \(b^{req}\), \(b^{res}\):} the size of incoming and outgoing data, capturing transfer-related effects.
  \item \textbf{HTTP method \(m\):} the request method (e.g., \texttt{GET}, \texttt{POST}, \texttt{PUT}, \texttt{DELETE}), which helps distinguish different execution paths and backend behaviours.
  \item \textbf{Response status \(r\):} captures outcomes such as success, redirection, rejection, or failure, which may correlate with different execution costs.
  \item \textbf{URL-parameter-derived features \(q\):} selected information derived from the request URI, such as the presence of specific query parameters, total query-string length, parameter count, or whitelisted numeric values (e.g., pagination limit). We use derived features rather than raw query strings in order to avoid unnecessary sparsity and high-cardinality model inputs.
  \item \textbf{Token count \(\tau\):} for AI inference, prompt and generated token counts exposed by the application.
\end{itemize}

The resulting dataset is a collection of tuples \((e, x, E)\) per endpoint \(e\), where \(x\) is the vector of relevant features. We then fit the constant, linear, and piecewise curve models described in Section~\ref{sec:methods}, and export the fitted parameters as will be described in the next section.

\section{Case Study: HTTP Server Implementation}
\label{sec:implementation}
We realise per-request estimation and disclosure as an nginx extension built on the \texttt{ngx\_http\_js\_module} (using the \texttt{njs} runtime), which avoids kernel modifications or external telemetry interfaces.

\subsection{Design and Runtime Behaviour}
At startup, the module loads the JSON registry described below and normalises it once per worker: routes are canonicalised, coefficients preloaded, and curve support points sorted. For each request, the module resolves the matching endpoint, optionally narrows the variant by HTTP method, extracts the required input features from variables already exposed by nginx (request time, request and response sizes, and, where applicable, application-supplied metadata such as token counts), and evaluates the model. Constant models return a fixed value; linear models apply preconfigured coefficients; curve models interpolate between precomputed support points. The energy estimate is then converted to operational and embodied emissions using the configured carbon parameters and exposed via response headers; values can optionally be written to structured logs for downstream analysis.

\subsection{Model Representation}
The JSON registry defines how endpoint-specific energy models and their parameters are represented for runtime evaluation. At the top level, the registry is a map from endpoint identifiers (e.g., URI paths or location blocks) to per-endpoint configuration objects. Each such object contains an \texttt{energy\_model} plus the shared parameters needed later for emission calculations, namely grid intensity and the embodied-allocation rate. The \texttt{energy\_model} itself is a tagged structure whose fields depend on the chosen model family: constant models store a single value, linear models store an intercept and coefficients, and curve models store support points together with interpolation rules. The registry format also allows multiple model variants for the same URL path, distinguished by an optional \texttt{filter} object. This is useful when, for example, \texttt{GET} and \texttt{POST} requests share the same endpoint path but exhibit different behaviour and should therefore be matched to different energy models. Listing~\ref{lst:json-registry} does not show the full generated registry, but a compact excerpt with representative instances of the model families used in the prototype.

\begin{lstlisting}[float,language=json,caption={Excerpt from the JSON model registry with representative endpoint entries},label={lst:json-registry}]
{
  "/ai": {
    "filter": { "method": "POST" },
    "energy_model": {
      "kind": "linear",
      "intercept_mWh": 7.062,
      "prompt_token_coeff_mWh_per_token": 0.017,
      "generated_token_coeff_mWh_per_token": 0.483
    },
    "embodied_rate_gCO2e_per_s": 0.001,
    "grid_intensity_gCO2e_per_kWh": 121.111
  },
  "/done": {
    "energy_model": {
      "kind": "constant",
      "value_mWh": 0.085
    },
    "embodied_rate_gCO2e_per_s": 0.001,
    "grid_intensity_gCO2e_per_kWh": 121.111
  },
  "/getToDos": {
    "energy_model": {
      "kind": "curve",
      "input": "data_size",
      "points": [[49421.59, 0.072], [231310.93, 0.112], [2051378.03, 0.125], [20298823.13, 0.553]],
      "interpolate": "linear",
      "extrapolate": "linear_tail"
    },
    "embodied_rate_gCO2e_per_s": 0.001,
    "grid_intensity_gCO2e_per_kWh": 121.111
  },
  ...
}
\end{lstlisting}

Energy is expressed in \(\mathrm{mWh}\); grid intensity in \(\mathrm{gCO_2e/kWh}\); embodied emissions as a time-based rate in \(\mathrm{gCO_2e/s}\), so per-request embodied impact follows from request duration. The grid-intensity field can be written statically or rewritten by a side process (e.g., fetching from Electricity Maps~\cite{electricitymaps} or WattTime~\cite{watttime}); the registry is reloaded on configuration reload, so no nginx restart is required.

\subsection{Disclosure Interface}
\label{subsec:interface}
The computed values are exposed via response headers, separating the underlying components of the calculation rather than encoding everything in a single aggregated value. This keeps the modelling assumptions explicit at the boundary:
\begin{itemize}
  \item \texttt{Request-Energy}: per-request energy, in \(\mathrm{mWh}\).
  \item \texttt{Grid-Intensity}: grid intensity in \(\mathrm{gCO_2e/kWh}\).
  \item \texttt{Request-Embodied-CO2e}: embodied share allocated to the request, in \(\mathrm{mgCO_2e}\).
  \item \texttt{Request-SCI}: total per-request emissions, in \(\mathrm{mgCO_2e}\).
\end{itemize}
These fields illustrate one possible design and are not proposed for standardisation. Listing~\ref{lst:curl} shows an example response with the four headers emitted (unrelated headers omitted for brevity).

\begin{lstlisting}[float,language=bash,caption={Example response headers emitted by the prototype nginx module},label={lst:curl}]
$ curl -sS -D - -o /dev/null \
  -X POST "http://localhost:8080/login" \
  -H "Content-Type: application/json" \
  -d "{\"username\":\"$USERNAME\",\"password\":\"$PASSWORD\"}"

HTTP/1.1 200 OK
Server: nginx/1.25.5
...
Set-Cookie: csrftoken=xxx; ...
Set-Cookie: sessionid=xxx; ..
(*@{\bfseries\ttfamily Request-Energy: 4.038}@*)
(*@{\bfseries\ttfamily Grid-Intensity: 121.111}@*)
(*@{\bfseries\ttfamily Request-Embodied-CO2e: 0.143}@*)
(*@{\bfseries\ttfamily Request-SCI: 0.633}@*)
\end{lstlisting}

\section{Evaluation}
\label{sec:evaluation}

We evaluate the approach as a feasibility study: the benchmark suite was executed ten times to stabilise the measurements, after which we inspect the derived model parameters, perform one trace-based replay validation against an independent click-through workload, and compare paired runs with and without header emission to quantify overhead. The results support claims about operational plausibility and order-of-magnitude agreement, not a full statistical characterisation of model error across deployments.

\subsection{Model Derivation and Parametrisation}

We convert the GMT export into the JSON registry of Listing~\ref{lst:json-registry}, instantiating all three model families. Family assignment was driven by workload-scaling benchmarks: for example, we issued between 1 and 1000 calls to \texttt{/login} to confirm that per-request energy remained stable~\cite{gmt-run-login-scaling}. \texttt{/login}, \texttt{/logout}, \texttt{/done}, and \texttt{/deleteAllToDos} produced near-constant profiles and use constant models;\\ \texttt{/createToDo} and \texttt{/getToDos} showed payload-dependent scaling and use curve models with measured support points and piecewise-linear interpolation; \texttt{/ai} uses a least-squares linear model over an intercept plus prompt- and generated-token counts.

For the measured system, grid intensity is configured statically at \(\num{121.111}\,\mathrm{gCO_2e/kWh}\) and the embodied allocation rate at \\ \(\num{0.001}\,\mathrm{gCO_2e/s}\); in production, the intensity could be refreshed from Electricity Maps~\cite{electricitymaps} or WattTime~\cite{watttime} (Section~\ref{subsec:emissions}). The registry is derived from total machine energy rather than idle-subtracted measurements, and HTTPS/TLS overhead is excluded from the calibration; both choices are limitations discussed in Section~\ref{sec:limitations}.

\begin{table}[h]
\small
\centering
\caption{Endpoint benchmark conditions and derived per-request energy labels.}
\label{tab:energy}
\begin{tabular}{p{4.3cm}rr}
\toprule
\textbf{Endpoint / condition} & \textbf{Requests} & \textbf{\makecell[r]{Energy\\ {\footnotesize (mWh/req)}}} \\
\midrule
\texttt{/login}                              & 100 & 4.038 \\
\texttt{/logout}                             & 100 & 0.006 \\
\texttt{/done}                               & 100 & 0.085 \\
\texttt{/deleteAllToDos}                     & 100 & 2.835 \\
\midrule
\texttt{/getToDos} (100 items)               & 100 & 0.072 \\
\texttt{/getToDos} (1\,000 items)            & 100 & 0.112 \\
\texttt{/getToDos} (10\,000 items)           & 100 & 0.125 \\
\texttt{/getToDos} (100\,000 items)          & 100 & 0.553 \\
\texttt{/createToDo} (text, 100 chars)       & 100 & 0.064 \\
\texttt{/createToDo} (text, 1\,000 chars)    & 100 & 0.079 \\
\texttt{/createToDo} (text, 10\,000 chars)   & 100 & 0.035 \\
\texttt{/createToDo} (text, 100\,000 chars)  & 100 & 0.096 \\
\texttt{/createToDo} (file, 1\,KB)           & 100 & 0.058 \\
\texttt{/createToDo} (file, 10\,KB)          & 100 & 0.061 \\
\texttt{/createToDo} (file, 100\,KB)         & 100 & 0.080 \\
\texttt{/createToDo} (file, 1\,MB)           & 100 & 0.288 \\
\texttt{/createToDo} (file, 5\,MB)           & 100 & 0.989 \\
\midrule
\texttt{/ai} (very short prompt)             &  10 & 29.978 \\
\texttt{/ai} (short prompt)                  &  10 & 251.044 \\
\texttt{/ai} (medium prompt)                 &  10 & 299.771 \\
\texttt{/ai} (long prompt)                   &  10 & 229.718 \\
\texttt{/ai} (very long prompt)              &  10 & 274.852 \\
\bottomrule
\end{tabular}
\end{table}

\subsection{Endpoint behaviour}
\begin{figure*}[t]
\centering
\begin{minipage}{0.46\textwidth}
\centering
\begin{tikzpicture}
\begin{axis}[
  width=\linewidth, height=5cm,
  xmode=log, log basis x={10},
  xlabel={request data size (bytes, log scale)},
  ylabel={energy per request (mWh)},
  title={\texttt{/createToDo}: curve model},
  title style={font=\small},
  label style={font=\small},
  tick label style={font=\footnotesize},
  grid=major, grid style={dashed,gray!30},
  legend pos=north west,
  legend style={font=\footnotesize,fill=white,fill opacity=0.9,draw=gray!50},
]
\addplot+[mark=*,mark size=2pt,blue,thick] coordinates {
  (100,0.064) (1000,0.079) (1124,0.058) (10000,0.035) (10340,0.061)
  (100000,0.096) (102500,0.080) (1048676,0.288) (5242980,0.989)
};
\addlegendentry{measured points + linear interpolation}
\end{axis}
\end{tikzpicture}
\end{minipage}\hfill
\begin{minipage}{0.46\textwidth}
\centering
\begin{tikzpicture}
\begin{axis}[
  width=\linewidth, height=5cm,
  xlabel={predicted energy per request (mWh)},
  ylabel={measured energy per request (mWh)},
  title={\texttt{/ai}: measured vs.\ predicted},
  title style={font=\small},
  label style={font=\small},
  tick label style={font=\footnotesize},
  grid=major, grid style={dashed,gray!30},
  xmin=0, xmax=320, ymin=0, ymax=320,
  legend pos=north west,
  legend style={font=\footnotesize,fill=white,fill opacity=0.9,draw=gray!50},
]
\addplot[no marks,dashed,black!50,domain=0:320,samples=2] {x};
\addlegendentry{$y=x$}
\addplot+[only marks,mark=o,mark size=2.5pt,blue,thick] coordinates {
  (29.813,29.978) (252.043,251.044) (299.070,299.771) (229.691,229.718) (274.619,274.852)
};
\addlegendentry{benchmark conditions}
\end{axis}
\end{tikzpicture}
\end{minipage}
\caption{Fitted models for one endpoint from each class.}
\label{fig:model-fit}
\end{figure*}
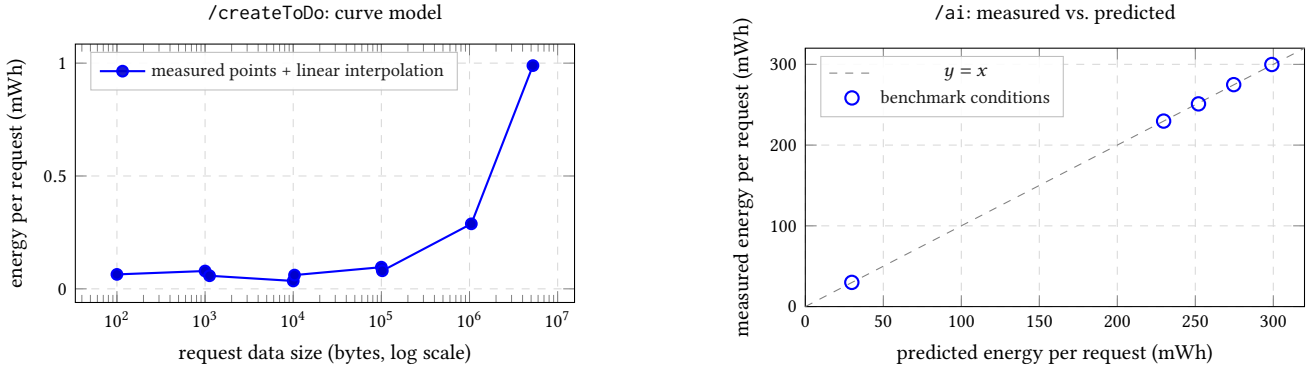

We evaluate the fitted models across the different endpoint classes in the benchmark to assess whether they preserve the expected workload-dependent behaviour. We distinguish endpoints by whether their dominant cost driver is visible at the HTTP boundary or only at the application layer: the conventional CRUD-style endpoints fall in the first group, while the inference endpoint serves as an example of the second. Figure~\ref{fig:model-fit} summarises the fitted models for one representative endpoint from each group.

\subsubsection{CRUD endpoints}
The generated registry shows that the REST endpoints fall into distinct behavioural classes. Authentication-related requests are close to constant cost. The fitted \texttt{/logout} model is only \SI{0.006}{mWh} per request, while \texttt{/login} is substantially larger at \SI{4.038}{mWh}, reflecting additional work such as credential verification and cryptographic session establishment. The \texttt{/done} endpoint is also effectively constant, with a fitted value of \SI{0.085}{mWh} per request across all measured conditions.

In contrast, the payload-dependent endpoints exhibit clear scaling behaviour. For \texttt{/createToDo}, the fitted curve (Figure~\ref{fig:model-fit}, left) remains nearly flat for small text-only requests but increases sharply once file uploads become large. A request with a short text body requires about \SI{0.064}{mWh}, while the \SI{5}{MB} attachment condition reaches \SI{0.989}{mWh}, corresponding to more than a fifteen-fold increase. Intermediate file sizes (\SI{1}{KB}, \SI{10}{KB}, \SI{100}{KB}, \SI{1}{MB}) produce a consistent progression, indicating that the curve model captures a meaningful size effect rather than measurement noise.

For \texttt{/getToDos}, the fitted model scales with the size of the returned collection. The smallest measured response proxy is approximately \SI{49.4}{kB} and corresponds to \SI{0.072}{mWh} per request, while the largest condition reaches about \SI{20.3}{MB} and \SI{0.553}{mWh}. This shows that response-size observables available at the HTTP boundary are sufficient to approximate workload-dependent behaviour without access to internal database operations.

These results illustrate that the fitted models capture the dominant differences between endpoint classes while remaining compact. The goal is not to reproduce the full internal behaviour of each endpoint, but to provide interface-level abstractions that preserve the major decision-relevant distinctions between small and large requests.

\subsubsection{Inference endpoint with application-supplied features}
The inference endpoint is used here as a test of whether the method extends to endpoints whose dominant cost driver is not visible at the HTTP boundary. Such endpoints require application-supplied features in addition to, or instead of, server-level observables; in this case, prompt-token and generated-token counts produced by the inference runtime. The observation that motivates this is that, when considering only prompt-profile labels (\texttt{very-short}, \texttt{short}, \texttt{medium}, \texttt{long}, \texttt{very-long}), the observed runtimes are not strictly monotonic. In our measurements, the \texttt{medium} profile took longer than the \texttt{very-long} profile, which appears inconsistent with the expectation that longer prompts should require more work.

This behaviour is explained by the underlying token statistics. While prompt-token counts increase monotonically across profiles (from roughly 20 to about 1010 tokens), generated-token counts vary substantially and do not follow the same ordering. In this measurement, average generated-token counts were approximately 46.4 (\texttt{very-short}), 505.8 (\texttt{short}), 600.7 (\texttt{medium}), 446.5 (\texttt{long}), and 518.4 (\texttt{very-long}). As a result, prompt categories alone are not reliable predictors of runtime or energy.

Using token-level information, the fitted model for \texttt{/ai} is given by
\[
\widehat{E}_{ai}(\tau^{prompt}, \tau^{gen}) =
7.062 + 0.017 \cdot \tau^{prompt} + 0.483 \cdot \tau^{gen}.
\]
The coefficients make the relative contribution of the two factors explicit. The weight assigned to generated tokens is roughly 28 times larger than that of prompt tokens, indicating that the dominant marginal work occurs during token generation rather than prompt processing. This is consistent with the behaviour of decoder-style inference systems. The right-hand panel of Figure~\ref{fig:model-fit} plots predicted against measured per-request energy for all five prompt profiles; predictions agree with measurements to within approximately \SI{1}{mWh} per condition, indicating that prompt-token and generated-token counts together explain almost all of the observed variation across the benchmarked workload.

This result demonstrates that the application-supplied feature mechanism is necessary rather than optional: for endpoints in this class, HTTP-level observables such as payload size or coarse prompt labels are insufficient to capture the dominant cost driver. The inference endpoint here is one example of this pattern; other endpoints whose work depends on application-internal variables (for example, result-set cardinality, cache state, or downstream service invocations) can be accommodated by the same mechanism. We make no broader claim about the carbon footprint of inference services at scale; the demonstration here is confined to the locally served endpoint in this case study.

\subsection{Click-Path Validation}
To assess how well the fitted endpoint models generalise beyond the isolated benchmark conditions used for calibration, we executed an additional no-header validation run that repeated a realistic user click-through path 100 times~\cite{gmt-run-validation-walkthrough}. Each iteration performed the sequence \texttt{/login} \(\rightarrow\) \texttt{/deleteAllToDos} \(\rightarrow\) \texttt{/createToDo} \(\rightarrow\) \texttt{/getToDos} \(\rightarrow\) \texttt{/done} \(\rightarrow\) \texttt{/getToDos} \(\rightarrow\) \texttt{/logout}, resulting in 700 HTTP requests in total. The total run was measured ten times with Green Metrics Tool to obtain the machine-level reference energy and, independently, nginx access logs were replayed offline through the derived registry to obtain the model-based estimate.

The GMT reported a total machine energy of
\SI{597.18}{mWh} (\(\pm\)\,\SI{0.74}{\percent}, average \(\pm\) standard deviation over all runs)~\cite{gmt-run-validation-reference}, while the offline log replay produced an estimated total of \SI{611.101}{mWh}. This corresponds to an absolute deviation of
\SI{13.921}{mWh}, or \SI{2.33}{\percent} relative to the measured total. For an
endpoint model derived from separate controlled measurements and then evaluated
only from runtime-visible log features, this level of agreement is encouraging:
the estimate preserves the correct order of magnitude and remains close enough
to support relative comparisons and aggregate observability.

\subsection{Nginx Plugin Overhead}
We evaluate runtime overhead using ten repeated-run comparisons with the nginx module disabled~\cite{gmt-run-overhead-disabled} and enabled~\cite{gmt-run-overhead-enabled}. Across the full run, mean machine energy rises from $5325.07 \pm 0.48\%$ to $5349.55 \pm 0.50\%\,\mathrm{mWh}$, an overhead of \SI{0.46}{\percent} (from the total-runtime entry of the GMT comparison export).

Per-route data is shown in Tables~\ref{tab:compare-constant-endpoints}--\ref{tab:compare-file-loads}. The differences are small and dominated by run-to-run variation. \texttt{/login}, the largest condition, changes by \SI{-0.3}{\percent}. \texttt{/logout} appears to rise \SI{10.9}{\percent}, but at \SI{15}{mWh} per condition with standard deviations near \SI{18}{\percent} this is better read as noise. For \texttt{/getToDos} and \texttt{/done} the changes range from roughly $-4$ to $+10$ percent across conditions; for \texttt{/createToDo} from $+0.1$ to $+11.1$ percent, with the largest relative differences on the smallest absolute conditions. The practically relevant \SI{1}{MB} and \SI{5}{MB} uploads change by \SI{0.2}{\percent} and \SI{0.1}{\percent}.

We do not tabulate the AI endpoint: across the five prompt profiles, mean energy differences remained within $\pm 1$ percent and runtime changes within $0.3$ percent, indistinguishable from inference variability. Overall, per-request model evaluation and disclosure can be integrated at the HTTP server with low overhead.

\begin{table}[h]
\centering
\small
\caption{Measured machine energy for authentication endpoints.}
\label{tab:compare-constant-endpoints}
\begin{tabular}{llrr}
\toprule
Class & Variant & \makecell[r]{Login\\(mWh)} & \makecell[r]{Logout\\(mWh)} \\
\midrule
Constant & No Header & 678.59 $\pm$ 0.88\% & 14.42 $\pm$ 18.44\% \\
 & Header & 676.31 $\pm$ 0.95\% & 15.99 $\pm$ 17.26\% \\
 & $\Delta$ & -0.3\% & +10.9\% \\
\bottomrule
\end{tabular}
\end{table}

\begin{table}[h]
\footnotesize
\centering
\caption{Measured machine energy for text-body workloads.}
\label{tab:compare-text-loads}
\setlength{\tabcolsep}{3pt}
\resizebox{\linewidth}{!}{%
\begin{tabular}{llrrr}
\toprule
\makecell[l]{Load\\(chars)} & Variant & \makecell[r]{CreateToDo\\(mWh)} & \makecell[r]{GetToDos\\(mWh)} & \makecell[r]{Done\\(mWh)} \\
\midrule
100 & No Header & 37.68 $\pm$ 5.74\% & 20.99 $\pm$ 5.49\% & 34.35 $\pm$ 14.57\% \\
 & Header & 39.33 $\pm$ 8.94\% & 20.21 $\pm$ 7.29\% & 37.82 $\pm$ 9.90\% \\
 & $\Delta$ & +4.4\% & -3.8\% & +10.1\% \\
\midrule
1000 & No Header & 39.63 $\pm$ 15.15\% & 21.31 $\pm$ 7.39\% & 34.95 $\pm$ 12.54\% \\
 & Header & 43.16 $\pm$ 10.45\% & 21.71 $\pm$ 6.67\% & 33.49 $\pm$ 13.69\% \\
 & $\Delta$ & +8.9\% & +1.9\% & -4.2\% \\
\midrule
10000 & No Header & 42.48 $\pm$ 10.15\% & 30.26 $\pm$ 4.15\% & 39.41 $\pm$ 9.45\% \\
 & Header & 44.20 $\pm$ 7.06\% & 31.14 $\pm$ 4.23\% & 38.84 $\pm$ 5.06\% \\
 & $\Delta$ & +4.0\% & +2.9\% & -1.5\% \\
\midrule
100000 & No Header & 46.05 $\pm$ 7.32\% & 131.92 $\pm$ 2.37\% & 45.58 $\pm$ 10.27\% \\
 & Header & 48.82 $\pm$ 7.45\% & 131.35 $\pm$ 1.90\% & 49.35 $\pm$ 8.72\% \\
 & $\Delta$ & +6.0\% & -0.4\% & +8.3\% \\
\bottomrule
\end{tabular}%
}
\end{table}

\begin{table}[h]
\small
\centering
\caption{Measured machine energy for file-upload workloads.}
\label{tab:compare-file-loads}
\begin{tabular}{llr}
\toprule
File size & Variant & \makecell[r]{CreateToDo\\(mWh)} \\
\midrule
1KB & No Header & 39.98 $\pm$ 10.67\% \\
 & Header & 41.42 $\pm$ 15.37\% \\
 & $\Delta$ & +3.6\% \\
\midrule
10KB & No Header & 36.27 $\pm$ 9.13\% \\
 & Header & 40.32 $\pm$ 10.89\% \\
 & $\Delta$ & +11.1\% \\
\midrule
100KB & No Header & 41.10 $\pm$ 10.31\% \\
 & Header & 44.73 $\pm$ 10.08\% \\
 & $\Delta$ & +8.8\% \\
\midrule
1MB & No Header & 68.66 $\pm$ 3.17\% \\
 & Header & 68.78 $\pm$ 3.64\% \\
 & $\Delta$ & +0.2\% \\
\midrule
5MB & No Header & 232.39 $\pm$ 0.80\% \\
 & Header & 232.62 $\pm$ 0.97\% \\
 & $\Delta$ & +0.1\% \\
\bottomrule
\end{tabular}
\end{table}

\section{Limitations}
\label{sec:limitations}

We group limitations into four classes.

\textbf{Modelling abstraction.}
Per-request estimates are interface-level abstractions, not exact physical attributions; they expose a consistent and decision-relevant signal rather than ground truth. The model reflects only the factors in the feature set, so unobserved effects (resource contention, background load, scheduling, co-tenancy, hardware-specific optimisations) can introduce systematic error. The current formulation does not emit uncertainty intervals; this limits the interpretation of absolute values but, as with latency and throughput, does not invalidate the signal for relative comparison.

\textbf{Deployment dependence.}
Estimates depend on hardware generation, power management, virtualization, co-located workloads, and regional grid conditions. The same logical endpoint may exhibit different physical footprints across environments, and prediction error grows when the deployment diverges from the calibration setting. Periodic recalibration is therefore necessary; automating it on software releases or hardware changes is left to future work.

\textbf{System boundary and attribution.}
A request may involve caches, databases, queues, or external services. The boundary here is the benchmarked deployment, which can include co-measured dependencies but not opaque third-party services. CDNs and other intermediaries shift where energy is expended, raising attribution questions (origin vs.\ edge). Whether reported values are marginal or include baseline must also be disclosed; the specific choices made here are stated in Section~\ref{sec:evaluation}.

\textbf{Prototype scope.}
HTTPS/TLS overhead is excluded from calibration; in deployments where HTTPS is mandatory it can be modelled as a separate constant or as a protocol-keyed variant. Family assignment is manual rather than automated. The nginx integration uses \texttt{ngx\_http\_js\_module}, which adds small but measurable interpretation overhead; a native C extension would improve this.

\section{Discussion}

Beyond external disclosure, the approach also supports internal provider-side analytics. Software producers typically see request counts, latency, and cost but not the energy or emissions of their own endpoints. By combining offline-derived per-endpoint models with routine HTTP logs, providers can estimate and aggregate energy or \(\mathrm{CO_2e}\) across endpoints, customers, or time windows, even when the underlying infrastructure is opaque.

The case for disclosure as a precondition for constraint rests on an asymmetry: in current HTTP systems, clients and operators see latency, cost, and throughput but not environmental cost. This systematically excludes ecological considerations from the concrete software decisions that determine aggregate consumption. Making per-request impact visible does not, by itself, reduce demand; but it removes a structural barrier, since one cannot budget, cap, or differentially price what one cannot measure at the relevant granularity.

The mechanism is therefore best understood as a substrate for constraint-led design traditions rather than a substitute for them. Sufficiency-oriented practice~\cite{pargman2014rethinking,nardi2018computingwithinlimits} requires that clients and providers can compare interactions and shift toward lower-impact alternatives, which presupposes a comparable per-request signal at the interface. Degrowth-oriented critiques of computing~\cite{becker2023insolvent} require that the ecological cost of an interaction is inescapable at the point of demand rather than externalised through cloud abstractions that present resources as effectively unlimited~\cite{mytton2020hiding}. Low-impact server designs in the tradition of the solar website~\cite{dedecker2018solar} require operating envelopes that the underlying system can both honour and communicate; per-request impact gives such systems a concrete signal to expose to readers and clients. In each case, the contribution here is not a substitute for these strategies but the visibility layer they presuppose.

Three risks attach to disclosure of this kind. First, per-request values may be used for greenwashing: a provider could disclose small per-request numbers while aggregate demand grows unchecked. Second, the mechanism can shift responsibility asymmetrically, framing reduction as a client-side concern (``reduce your prompt length'') while infrastructure growth remains unquestioned. Third, quantification itself can naturalise the assumption that environmental impact is acceptable so long as it is accounted for. These risks do not invalidate the approach but they constrain the claims that can responsibly be made on it; the contribution is a reproducible, inspectable method for making environmental implications visible at the interface where decisions are made, not metrological certainty.

\section{Conclusion}
This paper argued that the missing piece in HTTP sustainability disclosure is not header-field semantics but a practical method for generating credible per-request values. We showed that controlled machine-level measurements can be transformed into compact endpoint models, stored in a JSON registry, and evaluated online at the HTTP boundary by an nginx-based prototype. The prototype represents heterogeneous endpoint classes, including both conventional CRUD routes and an AI inference endpoint, with low overhead.

What this makes visible is a precondition, not an end in itself. Per-request impact is a substrate for constraint-led design traditions, such as sufficiency and sobriety~\cite{pargman2014rethinking,nardi2018computingwithinlimits}, degrowth-oriented critiques~\cite{becker2023insolvent}, and low-impact server designs~\cite{dedecker2018solar}. These traditions all assume that the ecological cost of an interaction is visible where demand is generated rather than externalised through cloud abstractions~\cite{mytton2020hiding}. Visibility also carries hazards: per-request figures can greenwash undiminished growth, can shift responsibility onto clients while leaving infrastructure expansion unquestioned, and can naturalise the assumption that impact is acceptable so long as it is accounted for. We therefore offer this method not as metrological certainty but as a reproducible, inspectable way to bring ecological cost into the decisions where it currently has can not be considered. A natural next step is to re-implement the runtime path as a native nginx C extension, minimizing the interpretation overhead.

\section*{Funding}
This work was funded by the Deutsche Bundesstiftung Umwelt (DBU).

\clearpage
\bibliographystyle{ACM-Reference-Format}
\bibliography{bib}

@misc{easytodo,
  author       = {{ribalba}},
  title        = {{easytodo}: A simple ToDo REST API benchmark application},
  year         = {2026},
  howpublished = {GitHub repository},
  url          = {https://github.com/ribalba/Limits-2026/tree/main/code/todoapp}
}

@misc{greenmetricstool,
  author       = {Tarara, Arne and Mateas, Dan and Hoffmann, Geerd-Dietger},
  title        = {Green Metrics Tool},
  year         = {2026},
  howpublished = {GitHub repository},
  url          = {https://github.com/green-coding-solutions/green-metrics-tool}
}

@misc{gsf-sci,
  author       = {{Green Software Foundation Standards Working Group}},
  title        = {Software Carbon Intensity (SCI) Specification},
  year         = {2024},
  howpublished = {Specification},
  note         = {Version 1.1.0; accessed 2026-04-01},
  url          = {https://sci.greensoftware.foundation/}
}

@misc{iso14040,
  author       = {{International Organization for Standardization}},
  title        = {{ISO} 14040:2006 Environmental management --- Life cycle assessment --- Principles and framework},
  year         = {2006},
  howpublished = {International Standard},
  note         = {Edition 2; official standard page accessed 2026-04-01},
  url          = {https://www.iso.org/standard/37456.html}
}

@misc{iso14044,
  author       = {{International Organization for Standardization}},
  title        = {{ISO} 14044:2006 Environmental management --- Life cycle assessment --- Requirements and guidelines},
  year         = {2006},
  howpublished = {International Standard},
  note         = {Edition 1; official standard page accessed 2026-04-01},
  url          = {https://www.iso.org/standard/38498.html}
}

@misc{ghgproduct,
  author       = {{World Resources Institute (WRI)} and {World Business Council for Sustainable Development (WBCSD)}},
  title        = {Product Life Cycle Accounting and Reporting Standard},
  year         = {2011},
  howpublished = {Standard},
  url          = {https://ghgprotocol.org/product-standard}
}

@techreport{martin-http-carbon-emissions-scope-2-00,
  author      = {Martin, Bertrand},
  title       = {{HTTP Response Header Field: Carbon-Emissions-Scope-2}},
  institution = {Internet Engineering Task Force},
  publisher   = {Internet Engineering Task Force},
  type        = {Internet-Draft},
  number      = {draft-martin-http-carbon-emissions-scope-2-00},
  year        = {2023},
  month       = apr,
  day         = {3},
  pagetotal   = {5},
  note        = {Work in Progress},
  url         = {https://datatracker.ietf.org/doc/draft-martin-http-carbon-emissions-scope-2/00/}
}

@misc{httpwg-admin-issue-52-carbon-emissions,
  author       = {Nottingham, Mark},
  title        = {Communicating carbon emissions},
  howpublished = {GitHub issue \#52 in httpwg/admin},
  year         = {2023},
  note         = {Opened 2023-04-12; accessed 2026-04-01},
  url          = {https://github.com/httpwg/admin/issues/52}
}

@misc{ietf-http-wg-mailinglist-carbon-header-thread,
  author       = {Pardue, Lucas},
  title        = {Re: Introducing a new HTTP response header for Carbon Emissions calculation},
  howpublished = {IETF HTTP Working Group mailing list archive},
  year         = {2023},
  note         = {Message dated 2023-04-11; part of the thread initiated by Bertrand Martin; accessed 2026-04-01},
  url          = {https://lists.w3.org/Archives/Public/ietf-http-wg/2023AprJun/0020.html}
}

@techreport{besleaga-green-sustainability-header-00,
  author      = {Besleaga, Andrei Nicolae},
  title       = {{The Sustainability HTTP Response Header Field}},
  institution = {Internet Engineering Task Force},
  publisher   = {Internet Engineering Task Force},
  type        = {Internet-Draft},
  number      = {draft-besleaga-green-sustainability-header-00},
  year        = {2026},
  month       = feb,
  day         = {27},
  pagetotal   = {6},
  note        = {Work in Progress},
  url         = {https://datatracker.ietf.org/doc/draft-besleaga-green-sustainability-header/00/}
}

@article{guldner2024gsmm,
  author  = {Guldner, Achim and Bender, Rabea and Calero, Coral and Fernando, Giovanni S. and Funke, Markus and Gr{\"o}ger, Jens and Hilty, Lorenz M. and H{\"o}rnschemeyer, Julian and Hoffmann, Geerd-Dietger and Junger, Dennis and Kennes, Tom and Kreten, Sandro and Lago, Patricia and Mai, Franziska and Malavolta, Ivano and Murach, Julien and Oberg{\"o}ker, Kira and Schmidt, Benno and Tarara, Arne and De~Veaugh-Geiss, Joseph P. and Westing, Max and Wohlgemuth, Volker and Naumann, Stefan},
  title   = {Development and evaluation of a reference measurement model for assessing the resource and energy efficiency of software products and components---{Green Software Measurement Model (GSMM)}},
  journal = {Future Generation Computer Systems},
  volume  = {155},
  pages   = {402--418},
  year    = {2024},
  doi     = {10.1016/j.future.2024.01.033}
}

@article{nardi2018computingwithinlimits,
  author  = {Nardi, Bonnie and Tomlinson, Bill and Patterson, Donald J. and Chen, Jay and Pargman, Daniel and Raghavan, Barath and Penzenstadler, Birgit},
  title   = {Computing Within Limits},
  journal = {Communications of the ACM},
  year    = {2018},
  month   = oct,
  url     = {https://cacm.acm.org/research/computing-within-limits/},
  note    = {Accessed 2026-04-01}
}

@article{sala2024greenwebmeter,
  author  = {Sala, Antonello and Barbetti, Lorenzo and Rosini, Andrea},
  title   = {Green Web Meter: Structuring and Implementing a Real-Time Digital Sustainability Monitoring System},
  journal = {Sustainability},
  volume  = {16},
  number  = {17},
  pages   = {7627},
  year    = {2024},
  doi     = {10.3390/su16177627},
  url     = {https://doi.org/10.3390/su16177627}
}

@article{desislavov2023trends,
  author  = {Desislavov, Radosvet and Mart{\'i}nez-Plumed, Fernando and Hern{\'a}ndez-Orallo, Jos{\'e}},
  title   = {Trends in {AI} inference energy consumption: Beyond the performance-vs-parameter laws of deep learning},
  journal = {Sustainable Computing: Informatics and Systems},
  volume  = {38},
  pages   = {100857},
  year    = {2023},
  doi     = {10.1016/j.suscom.2023.100857}
}

@inproceedings{poddar2025towards,
  author    = {Poddar, Soham and Koley, Paramita and Misra, Janardan and Ganguly, Niloy and Ghosh, Saptarshi},
  title     = {Towards Sustainable {NLP}: Insights from Benchmarking Inference Energy in Large Language Models},
  booktitle = {Proceedings of the 2025 Conference of the Nations of the Americas Chapter of the Association for Computational Linguistics: Human Language Technologies (Volume 1: Long Papers)},
  year      = {2025},
  pages     = {12688--12704},
  address   = {Albuquerque, New Mexico},
  publisher = {Association for Computational Linguistics},
  doi       = {10.18653/v1/2025.naacl-long.632},
  url       = {https://aclanthology.org/2025.naacl-long.632/}
}

@article{yang2024queueing,
  author  = {Yang, Yuqing and Xu, Yuedong and Jiao, Lei},
  title   = {A Queueing Theoretic Perspective on Low-Latency {LLM} Inference with Variable Token Length},
  journal = {arXiv preprint arXiv:2407.05347},
  year    = {2024},
  doi     = {10.48550/arXiv.2407.05347},
  url     = {https://arxiv.org/abs/2407.05347}
}

@article{caravaca2025promptstopower,
  author  = {Caravaca, Francisco and Cuevas, {\'A}ngel and Cuevas, Rub{\'e}n},
  title   = {From Prompts to Power: Measuring the Energy Footprint of {LLM} Inference},
  journal = {arXiv preprint arXiv:2511.05597},
  year    = {2025},
  doi     = {10.48550/arXiv.2511.05597},
  url     = {https://arxiv.org/abs/2511.05597}
}

@article{parssinen2018onlineadvertising,
  author  = {P{\"a}rssinen, Matti and Kotila, Marko and Cuevas, Rub{\'e}n and Phansalkar, Akshay and Manner, Jukka},
  title   = {Environmental impact assessment of online advertising},
  journal = {Environmental Impact Assessment Review},
  volume  = {73},
  pages   = {177--200},
  year    = {2018},
  issn    = {0195-9255},
  doi     = {10.1016/j.eiar.2018.08.004}
}

@inproceedings{han2025fairco2,
  author    = {Han, Leo and Kakadia, Jash and Lee, Benjamin C. and Gupta, Udit},
  title     = {Fair-CO2: Fair Attribution for Cloud Carbon Emissions},
  booktitle = {Proceedings of the 52nd Annual International Symposium on Computer Architecture (ISCA '25)},
  year      = {2025},
  pages     = {646--663},
  publisher = {Association for Computing Machinery},
  address   = {New York, NY, USA},
  doi       = {10.1145/3695053.3731023}
}

@misc{greencoding_bestpractices,
  author       = {{Green Coding Solutions}},
  title        = {Best Practices -- Green Metrics Tool Documentation},
  year         = {2026},
  howpublished = {Online documentation},
  note         = {Accessed 2026-04-01},
  url          = {https://docs.green-coding.io/docs/measuring/best-practices/}
}

@article{mytton2020hiding,
  author  = {Mytton, David},
  title   = {Hiding greenhouse gas emissions in the cloud},
  journal = {Nature Climate Change},
  volume  = {10},
  number  = {8},
  pages   = {701},
  year    = {2020},
  doi     = {10.1038/s41558-020-0837-6}
}

@inproceedings{pargman2014rethinking,
  author    = {Pargman, Daniel and Raghavan, Barath},
  title     = {Rethinking sustainability in computing: From buzzword to non-negotiable limit},
  booktitle = {Proceedings of the 8th Nordic Conference on Human-Computer Interaction (NordiCHI '14)},
  year      = {2014},
  publisher = {Association for Computing Machinery},
  pages     = {638--647},
  doi       = {10.1145/2639189.2639228}
}

@book{becker2023insolvent,
  author    = {Becker, Christoph},
  title     = {Insolvent: How to Reorient Computing for Just Sustainability},
  publisher = {MIT Press},
  year      = {2023}
}

@inproceedings{raghavan2011energy,
  author    = {Raghavan, Barath and Ma, Justin},
  title     = {The energy and emergy of the Internet},
  booktitle = {Proceedings of the 10th ACM Workshop on Hot Topics in Networks (HotNets-X)},
  year      = {2011},
  publisher = {Association for Computing Machinery},
  doi       = {10.1145/2070562.2070571}
}

@misc{dedecker2018solar,
  author       = {De Decker, Kris},
  title        = {How to build a low-tech website?},
  year         = {2018},
  howpublished = {Low-tech Magazine},
  url          = {https://solar.lowtechmagazine.com/2018/09/how-to-build-a-low-tech-website/},
  note         = {Accessed 2026-05-25}
}

@misc{electricitymaps,
  author       = {{Electricity Maps}},
  title        = {Electricity Maps: Live and historical electricity carbon intensity by region},
  howpublished = {Online service},
  url          = {https://www.electricitymaps.com/},
  note         = {Accessed 2026-05-25}
}

@misc{watttime,
  author       = {{WattTime}},
  title        = {{WattTime}: Marginal emissions and grid carbon intensity data},
  howpublished = {Online service},
  url          = {https://www.watttime.org/},
  note         = {Accessed 2026-05-25}
}

@misc{gmt-run-benchmark-energy,
  author       = {Hoffmann, Geerd-Dietger},
  title        = {Green Metrics Tool run comparison: per-endpoint benchmarking measurements for per-request energy labels},
  year         = {2026},
  howpublished = {Green Metrics Tool measurement run},
  url          = {https://metrics.green-coding.io/compare.html?ids=4d0b0afa-a994-446e-b10c-a9afdcd0cfb4,22947267-6714-4fd2-9261-e86f2e17dea6,69192a65-366f-4298-ae44-2a3dbd1913d2,b73289cb-c090-4190-ab2e-223e676452db,ada88f63-0e14-4947-8b60-ea172dab3aa2,0c20ec45-3c70-4444-b75f-3a5ee3bd679d,813bbcab-8a24-4209-8ec4-d748561eba75,2b8cb442-f989-49b4-ba6a-bfa6b2094c38,b08d8b4d-c2a1-4c72-8033-eea52b9fb3a6,aac19717-b9b9-4454-a75c-89ff6e67d0d4},
  note         = {Accessed 2026-05-25}
}

@misc{gmt-run-login-scaling,
  author       = {Hoffmann, Geerd-Dietger},
  title        = {Green Metrics Tool run: \texttt{/login} workload-scaling benchmark (1--1000 calls)},
  year         = {2026},
  howpublished = {Green Metrics Tool measurement run},
  url          = {https://metrics.green-coding.io/stats.html?id=6c40a5b8-79b7-4372-a1c4-3ae7185451d2},
  note         = {Accessed 2026-05-25}
}

@misc{gmt-run-validation-walkthrough,
  author       = {Hoffmann, Geerd-Dietger},
  title        = {Green Metrics Tool run: no-header user click-through validation walkthrough},
  year         = {2026},
  howpublished = {Green Metrics Tool measurement run},
  url          = {https://metrics.green-coding.io/stats.html?id=fca8c396-7591-4f16-9fd5-c3f87c2a424f},
  note         = {Accessed 2026-05-25}
}

@misc{gmt-run-validation-reference,
  author       = {Hoffmann, Geerd-Dietger},
  title        = {Green Metrics Tool run comparison: ten-run reference energy for click-through validation},
  year         = {2026},
  howpublished = {Green Metrics Tool measurement run},
  url          = {https://metrics.green-coding.io/compare.html?ids=30ed795e-04ff-4555-82a3-863cbbdac0c3,072645a6-0dac-4952-a7f1-0f191cfbfa6e,17fee139-3d17-4610-8d46-449e7c500914,64c35e72-d728-42cb-a342-3e7807a6c006,9798dbb1-327e-4d9c-91e5-69719e021ca4,224b368c-63f0-4db8-9df0-d631e90ac6b8,fc6ed1c7-aff7-4fee-9bb7-7c2a917324c1,afc9898c-ec48-468b-8ddb-9a99def2d35a,691ccd74-fcd4-40c1-9781-740887661da0,a7642738-4bfa-4f42-b1be-06120e909f3f},
  note         = {Accessed 2026-05-25}
}

@misc{gmt-run-overhead-disabled,
  author       = {Hoffmann, Geerd-Dietger},
  title        = {Green Metrics Tool run comparison: runtime overhead measurements with nginx module disabled},
  year         = {2026},
  howpublished = {Green Metrics Tool measurement run},
  url          = {https://metrics.green-coding.io/compare.html?ids=87d1724b-81ec-44d7-a400-444f663b76e3,4d0b0afa-a994-446e-b10c-a9afdcd0cfb4,22947267-6714-4fd2-9261-e86f2e17dea6,69192a65-366f-4298-ae44-2a3dbd1913d2,b73289cb-c090-4190-ab2e-223e676452db,ada88f63-0e14-4947-8b60-ea172dab3aa2,0c20ec45-3c70-4444-b75f-3a5ee3bd679d,813bbcab-8a24-4209-8ec4-d748561eba75,2b8cb442-f989-49b4-ba6a-bfa6b2094c38,b08d8b4d-c2a1-4c72-8033-eea52b9fb3a6,aac19717-b9b9-4454-a75c-89ff6e67d0d4},
  note         = {Accessed 2026-05-25}
}

@misc{gmt-run-overhead-enabled,
  author       = {Hoffmann, Geerd-Dietger},
  title        = {Green Metrics Tool run comparison: runtime overhead measurements with nginx module enabled},
  year         = {2026},
  howpublished = {Green Metrics Tool measurement run},
  url          = {https://metrics.green-coding.io/compare.html?ids=37223dd2-9481-4009-a9cc-25944c40f677,328f8dc3-b7ed-44be-91d6-6db2a8962d88,c243bddb-3224-400e-b77f-5099c6bc2b1b,0c64123d-3649-47b7-aeb3-4f7d84eff32f,af8cbea1-e8ef-4dad-ba49-ec3e4f925689,cfbe8885-2498-4a2d-8dfd-dbc79a9a6910,41a0ad09-0891-4039-9ea6-973f466b5d2d,fda1a216-97c3-42b4-b747-9e6f478c2f5c,7d68c17b-1f25-4ad3-b1ce-37ab292418bb,7d5218ac-0d3b-42d1-8bca-80df094aedd8},
  note         = {Accessed 2026-05-25}
}

\end{document}